\documentclass[12pt,showpacs,preprintnumbers,amsmath,aps,amssymb]{revtex4}
\usepackage{graphicx}% Include figure files
\usepackage{dcolumn}% Align table columns on decimal point
\usepackage{bm}% bold math
\usepackage{color}
\begin{document}

\baselineskip 18pt

\title{Exact solutions using power law scalar potential in the Saez-Ballester-K-essence like theory }

\author{J. Socorro}
\email{socorro@fisica.ugto.mx}
 \affiliation{Departamento de
F\'{\i}sica, DCeI, Universidad de Guanajuato-Campus Le\'on, C.P.
37150, Le\'on, Guanajuato, M\'exico}
\author{A. Gil-Ocaranza}
\email{victor.gil@academicos.udg.mx} \affiliation{Departamento de
Física, Centro de Investigación y de Estudios Avanzados del
Instituto Politécnico Nacional Apartado Postal 14-740, 07000,
Ciudad de México, México.} \affiliation{Centro Universitario de la
Costa - UDG Av. Universidad de Guadalajara 203, Ixtapa, Los
Tamarindos, 48280 Puerto Vallarta, Jal.,México.}
\author{Ximena L\'opez-Mujica}
\email{x.lopezmujica@ugto.mx}
 \affiliation{Departamento de
F\'{\i}sica, DCeI, Universidad de Guanajuato-Campus Le\'on, C.P.
37150, Le\'on, Guanajuato, M\'exico}
\author{Cesar Aar\'on Pacheco-V\'azquez}
\email{ca.pachecovazquez@ugto.mx}
 \affiliation{Departamento de
F\'{\i}sica, DCeI, Universidad de Guanajuato-Campus Le\'on, C.P.
37150, Le\'on, Guanajuato, M\'exico}

\begin{abstract}
We investigate a K-essence like cosmological model whose
scalar-field potential is constructed from a negative power-law
S\'aez--Ballester potential. By means of a suitable field
redefinition from $\phi$ to $\varphi$, we show that the resulting
field equations acquire a mathematical structure analogous to that
of a previously solved Friedmann-Lema$\hat{\i}$tre-Robertson-Walker
(FLRW) cosmological model. This correspondence allows us to obtain
exact classical solutions for both the scale factor and the scalar
field within the Hamiltonian formalism. The resulting cosmological
dynamics exhibits a late-time accelerated expansion, with the
deceleration parameter approaching the asymptotic value
$q\rightarrow -1$, characteristic of a de Sitter phase. At the
quantum level, the corresponding Wheeler-DeWitt (WDW) equation is
derived and exact quantum solutions are obtained. These results
provide a consistent classical and quantum description of the
cosmological evolution generated by this class of K-essence models.
In this formalism, the scalar field remains as a cosmic background
where the universe unfolds, which is glimpsed from the quantum
solution perspective.

\end{abstract}
\maketitle
\section{Introduction}
Found exact classical solutions using power law scalar potential
$V(\phi)=V_0 \phi^{\pm \lambda}$ where $\lambda$ is a positive
constant, in quintessence and  phantom theories where the scalar
fields are present, these do not exist in literature, only very
specific solutions \cite{PRD-1999,EPJC-2014}, or just a treatment
from the point of view of dynamic systems, a method very useful in
these and other processes \cite{CQG-2014,PRD-2017,PDU-2024} and
references therein.

In the literature appear other formalism where the scalar fields are
introduced with some particularities, such is the case of the
K-essence models which are restricted to the Lagrangian density of
the form
\cite{IJTP-2014,roland,chiba,bose,arroja,tejeiro,Pereira-2024,Pereira-2025}
\begin{equation}
 S_K=\int d^4x \, \sqrt{-g}\,\left[ f(\phi) \, {\cal G}(X)\right],
\end{equation}
where the canonical kinetic energy is given by ${\cal
G}(X)=X=-\frac{1}{2}\nabla_\mu \phi \nabla^\mu \phi$, $f(\phi)$ is
an arbitrary function of the scalar field $\phi$, and g is the
determinant of the metric. Other theory is called Saez-Ballester
formalism which are restricted a particular way of K-essence-like
Lagrangian density of the form \cite{saez-ballester-1986}
\begin{equation}
 S_{S-B}=\int d^4x \, \sqrt{-g}\,\left[ \phi^m \, X\right],
\end{equation}
with $m \in \mathbb{R}$, but it is restricted to the value m=-2,
which is not allowed within its model, and take the form as in
K-essence formalism. For this formalism, however, in the literature
there are exact solutions for the cases $f(\phi)=\phi^m$ and
$f(\phi)=e^{m\phi}$, were solve in the classical and quantum regime
for the FRW and Bianchi Class A cosmological model
\cite{fizika-2010} and in anisotropic cosmological model
\cite{rmf-2010}. There are other one, as in multifield formalism,
where appear a lagrangian density in the scalar field
\begin{equation}
 S=\int d^4x \, \sqrt{-g}\,\left[ -\frac{1}{2}\nabla_\mu \phi \nabla^\mu \phi -\frac{1}{2}g^{\mu \nu} F(\phi) \nabla_\mu \psi \nabla_\nu \psi -V(\phi)\right],
\end{equation}
where the function $F(\phi)$ usually take the form as an
exponential, $e^{\pm \kappa \phi}$ \cite{CQG-2021}, $Sinh^2(\lambda
\phi)$, \cite{coupled}, or a family of them
\cite{PRD-2004-chimento}, in scalar-tensor theory
\cite{Elizalde-2004} or in generalized theory of a scalar field
coupled nonminimally to the curvature and to a Brans-Dicke-like
theory \cite{Elizalde-2008}.

 In this work we introduce a mixed formalism between K-essence and
 Saez-Ballester theory plus a potential scalar field, in the particular value in the m constant and
 using a power law in the scalar potential term $V(\phi)=V_0 \phi^{\pm
 \lambda}$, being
 our action as
\begin{equation}
 S_{SBK}=\int d^4x \, \sqrt{-g}\,\left[ G\left({\cal X}\right)-V_0 \phi^{\pm \lambda}\right],
\end{equation}
where now ${\cal X}=\epsilon \phi^{-2} \, X$, the parameter
$\epsilon$ take the values $+1$ for quintessence and $-1$ for
phantom fields, and making the transformation $\phi=e^{\varphi}$ we
obtain that ${\cal X}=-\epsilon \frac{1}{2}\nabla_\mu \varphi
\nabla^\mu \varphi$ and $V(\phi)\to V(\varphi)=V_0 e^{\pm \lambda
\varphi}$, in other words, it is transformed into a known problem
where exact solutions exist, the power law scalar potential this
problem is translated into a known one in another field $\varphi$ as
a exponential scalar field potential $V(\varphi)=V_0 e^{\pm \lambda
\varphi}$, for which there are exact solutions, both classical and
quantum behavior, also in scalar-tensor theory \cite{Elizalde-2004},
or its generalization \cite{Elizalde-2008}.

\subsection{Saez-Ballester-K-essence like theory   }
One of the  simplest Lagrangian density is
\begin{equation}
 {\cal L}_{geo}=\left( R+   {\cal G}({\cal X})+V(\varphi)\right),
\label{lagra}
\end{equation}
where $R$ is the scalar of curvature, $V(\phi)=V_0 e^{\pm \lambda
\varphi}$,  ${\cal G}({\cal X})$ are defined before in the $\varphi$
field. Then, the field equations are given the corresponding
variation of  (\ref{lagra}), with respect to the metric and the
scalar field $\varphi$, gives the Einstein  and Klein-Gordon field
equations
\begin{eqnarray}
&&\rm R_{\alpha\beta}-\frac{1}{2}g_{\alpha\beta}R=-\epsilon
\frac{1}{2}\left(\nabla_\alpha\varphi\nabla_\beta\varphi-\frac{1}{2}g_{\alpha\beta}
g_{\mu\nu}\nabla_\mu\phi\nabla_\nu\varphi\right)+\frac{1}{2}g_{\alpha\beta}V(\varphi), \label{camrel}\\
&&\rm \square\varphi-\frac{\partial
V}{\partial\varphi}=0.\label{klein}
\end{eqnarray}
From  (\ref{camrel}) it  can be deduced  that the energy-momentum
tensor associated with the scalar field is
\begin{equation}
\rm 8 \pi G
T^{(\varphi)}_{\alpha\beta}=\frac{\epsilon}{2}\left(\nabla_\alpha\varphi\nabla_\beta\varphi-\frac{1}{2}g_{\alpha\beta}
g_{\mu\nu}\nabla_\mu\varphi\nabla_\nu\varphi\right)-\frac{1}{2}g_{\alpha\beta}V(\varphi)
\end{equation}\\
 The line element to be considered in this work is
the flat Friedmann-Lema$\hat{\i}$tre-Robertson-Walker (FLRW)
\begin{eqnarray}
\rm ds^2&=&-N(t)^2 dt^2 +e^{2\Omega(t)} \left[dr^2
+r^2(d\theta^2+sin^2\theta d\phi^2) \right], \nonumber\\
&=&-d\tau^2 + e^{2\Omega(\tau)} \left[dr^2
+r^2(d\theta^2+sin^2\theta d\phi^2) \right], \label{frw}
\end{eqnarray}
where we identify the time transformation $\rm N(t) dt=d\tau$, this
transformation will be used in the whole work, and in special gauge
we recover directly the cosmic time t, where the scale factor $\rm
A(t)=e^{\Omega(t)}$ in the Misner's parametrisation. One way to
visualize inflationary behavior is by calculating the deceleration
parameter q, defined as
\begin{equation}
\rm q(t)=-\frac{A\ddot{A}}{\dot A^2}, \label{q-deceleration}
\end{equation}
with the deceleration parameter approaching the asymptotic value
$q\rightarrow -1$, characteristic of a de Sitter phase.

\subsection{field equations}
 Making use to the metric (\ref{frw}) and a comoving  fluid, the equations (\ref{camrel}) y (\ref{klein})
 becomes (a dot mean a time derivative)
\begin{eqnarray}\rm
\frac{3\dot{\Omega}^2}{N^2}-\epsilon\frac{\dot{\varphi}^2}{4N^2}-\frac{1}{2} V(\varphi)&=&0,\label{ein0}\\
\rm
\frac{2\ddot{\Omega}}{N^2}+3\frac{\dot{\Omega}^2}{N^2}-\frac{2\dot{\Omega}\dot{N}}{N^3}+\epsilon
\frac{ \dot{\varphi}^2}{4N^2}
-\frac{1}{2}V(\varphi)&=&0,\label{eini}\\
\rm \left[-3{\dot \Omega}\frac{{\dot \varphi}}{N^2}-\frac{\ddot
\varphi}{N^2}+\frac{\dot \varphi}{N}\frac{\dot
N}{N^2}\right]-\frac{\partial V}{\partial \varphi}&=&0 \label{kg},
\end{eqnarray}
using the time transformation $\rm d\tau=Ndt$, and the chain rule
$\frac{\partial V}{\partial\varphi}=\frac{\partial V}{\partial
\tau}\frac{\partial \tau}{\partial\varphi}=
\frac{V^\prime}{\varphi^\prime}$ in the last equation we obtain

\begin{eqnarray}\rm 3{\Omega}^{\prime 2}-\epsilon \frac{{\varphi}^{\prime 2}}{4}-\frac{1}{2}V(\varphi)&=&0,\label{ein0-t}\\
\rm 2{\Omega}^{\prime \prime}+3{\Omega}^{\prime 2}+\epsilon \frac{
{\varphi}^{\prime 2}}{4}
-\frac{1}{2}V(\varphi)&=&0,\label{eini-t}\\
\rm \epsilon \left[-3{ \Omega^{\prime}}{\varphi^{\prime 2}}-
\varphi^{\prime \prime}\varphi^{\prime}\right]-
V^{\prime}=&0,\label{kgord}
\end{eqnarray}
the Klein-Gordon equation (\ref{kgord}) can be rewritten as
\begin{equation}
\rm \frac{d}{d\tau}\left[Ln\left(e^{6\Omega} \epsilon
\frac{\varphi^{\prime 2}}{2}\right)
\right]=-\frac{V^\prime}{\frac{\varphi^{\prime 2}}{2}},\label{kg0}
\end{equation}
Since a power law is proposed for scalar potentials with both signs,
however  to illustrate the methodology, we will specialize only in
the case of negative powers in the scalar field.

\section{Negative power law $V(\phi)=\phi^{-\lambda} \to\,
V(\varphi)=V_0 e^{-\lambda \varphi}$}

In this approach we use the particular scalar field potential $\rm
V(\varphi)=V_0 e^{-\lambda \varphi}$, when the parameter take the
particular value $\lambda=\sqrt{3}$, this class of potential  appear
in the supersymmetric quantum mechanics as the appropriate to have a
super inflation evolution in the scale factor of the universe
\cite{sodo,nuevo}, so (\ref{kg0}) is written as
\begin{equation}
\rm \frac{d}{d\tau}\left[Ln\left(e^{6\Omega}\epsilon
\frac{\varphi^{\prime 2}}{2}\right) \right]=\frac{2\lambda
V(\tau)}{\varphi^{\prime}},\label{kg-lambda}
\end{equation}

The question is, which are the time dependence solutions for these
$\rm (\Omega,\phi)$ functions?. The algebraic structure of the EKG
does not permit solver these one, then is necessary use other method
for solve. In the following section se implement the the Hamilton
approach, for obtain the exact solutions of these equation, without
use any ansatz for the scale factor like $\rm \Omega$ and the scalar
field $\rm \varphi$ .

\section{Lagrangian and Hamiltonian density \label{ham}}
 For obtain the classical solution to
Einstein-Klein-Gordon equations (\ref{camrel}) and (\ref{klein}) we
shall use the Hamilton approach, so we need to built the
corresponding Lagrantian and Hamiltoniad density for this
cosmological model.

In this way, we use (\ref{frw}) into (\ref{lagra}) we have
\begin{equation}
\rm {\cal L}=e^{3\Omega}\left[6\frac{{\dot \Omega}^2}{N}-\epsilon
\frac{1}{2}\frac{\dot \varphi^2}{N}+ N V_0 e^{-\lambda
\varphi}\right], \label{lagrafrw}
\end{equation}
the corresponding momenta are defined in the usual way $\rm
\Pi_q=\frac{\partial {\cal L}}{\partial \dot q}$,
\begin{eqnarray}
\rm \Pi_\Omega&=& \rm 12 \frac{e^{3\Omega}}{N}\dot \Omega,
\qquad\qquad \dot
\Omega=\frac{N e^{-3\Omega}}{12}\Pi_\Omega, \nonumber\\
\rm \Pi_\varphi&=&\rm -\epsilon \frac{e^{3\Omega}}{N}\dot
\varphi,\qquad\qquad \,\, \dot \varphi=-\frac{N}{\epsilon}
e^{-3\Omega} \Pi_\varphi. \label{momenta}
\end{eqnarray}
and the Hamiltonian density written as $\rm {\cal L}=\Pi_q \dot
q-N{\cal H}$, when we perform the variation of this canonical
lagrangian with respect to N, $\frac{\delta {\cal
L}_{canonical}}{\delta N} =0$, implying the constraint ${\cal
H}_I=0$. In our model the only constraint corresponds to Hamiltonian
density, which is weakly zero.
\begin{equation}
\rm {\cal H}= \frac{e^{-3\Omega}}{24} \left[
\Pi_\Omega^2-\frac{12}{\epsilon} \Pi_\varphi^2-24  V_0
e^{6\Omega-\lambda \varphi}\right]. \label{hamifrw}
\end{equation}
In the gauge $\rm N=24 e^{3\Omega}$ and using the Hamilton approach,
we have the following set of equation
\begin{eqnarray}
\rm \dot \Omega&=&\rm 2 \Pi_\Omega, \qquad \dot \varphi= -\frac{24}{\epsilon} \Pi_\varphi, \nonumber\\
\rm \dot \Pi_\Omega&=& \rm 144 V_0 e^{6\Omega- \lambda \varphi},
\qquad \dot \Pi_\phi= -24\lambda V_0 e^{6\Omega- \lambda
\varphi},\label{new-variables}
\end{eqnarray}

from (\ref{new-variables}) we find the relation between the momenta
\begin{equation}
\Pi_\varphi=-\frac{\lambda\Pi_\Omega}{6}+p_0 \label{ecPiphi}
\end{equation}
that inserting in the equation for the scalar field
\begin{equation}
\varphi=2\lambda\Omega-24p_0t+\alpha_0. \label{ec-phi}
\end{equation}

Employing the hamiltonian constraint, we find the following master
equation
\begin{equation}
\frac{d\Pi_\Omega}{a\Pi^2_\Omega +b\Pi_\Omega-c}=dt, \label{master}
\end{equation}
with the constants $a=2(3-\frac{\lambda^2}{\epsilon})$,
$b=\frac{24\lambda p_0}{\epsilon}$ and $c=\frac{72p_0^2}{\epsilon}$.

At this point we will make the separation to study two possible
cases of scalar fields: quintessence or phantom, depending of the
sign in the parameters $\epsilon$, first we considering  for
quintessence we have $\epsilon=1$, then we have

%%%%%%%%
%%%%%%%%
\subsection{Quintessence scalar field, $\epsilon=1$}
For this case, in the equation (\ref{master}) the parameter are
$a=2(3-\lambda^2)$, $b=24\lambda p_0$ and $c=72p_0^2$, and depending
to the value in the parameter $a$ there are different solutions
\begin{enumerate}
\item{} $a=0$, imply that $\lambda=\sqrt{3}$ obtaining the solution
for the momenta $\Pi_\Omega$ as
\begin{equation}
\Pi_\Omega= \sqrt{3}p_0+\frac{\sqrt{3}p_1}{72p_0} e^{24\sqrt{3}p_0
t}, \label{p1}
\end{equation}
so, the momenta
$\Pi_\varphi=\frac{p_0}{2}-\frac{p_1}{144p_0}e^{24\sqrt{3}p_0 t}$
and the functions $(\Omega,\varphi)$, from (\ref{new-variables}),
are
\begin{equation}
\Omega(t)=\Omega_0 +2\sqrt{3}p_0
t+\frac{p_1}{864p_0^2}e^{24\sqrt{3}p_0\,t}, \qquad
\varphi=\varphi_0-12p_0\,t+\frac{\sqrt{3}p_1}{432p_0^2}e^{24\sqrt{3}p_0
t}, \label{sols}
\end{equation}
and using the equation (\ref{ec-phi}) we find the relation between
the constants $\varphi_0=2\sqrt{3}\Omega_0 + \alpha_0$.

Thus, the scale factor A and the real scalar field $\phi$ become
\begin{eqnarray}
A(t)&=&A_0 Exp\left[2\sqrt{3}p_0\, \Delta
t+\frac{p_1}{864p_0^2}e^{24\sqrt{3}p_0\,t} \right],
\label{scale-factor}\\
\phi(t)&=&\phi_0\,Exp\left[-12p_0\,t+\frac{\sqrt{3}p_1}{432p_0^2}e^{24\sqrt{3}p_0
t} \right]. \label{scalar-0}
\end{eqnarray}
 The exact solutions show that the model describes a continuously
expanding universe, with the volume function increasing monotonically
throughout the evolution. At the same time, the scalar field displays a
smooth but nontrivial dynamics, reflecting the influence of the inverse
power-law potential. For the representative choice
$\lambda=\sqrt{3}$, corresponding to
$V(\phi)=V_0\phi^{-\sqrt{3}}$, the scalar field remains dynamically
active during the entire evolution, while the expansion is driven by
the steadily increasing volume function, as shown in
Fig.~\ref{volumen-a0}.

\begin{figure}[h!]
\begin{center}
\includegraphics[scale=0.5]{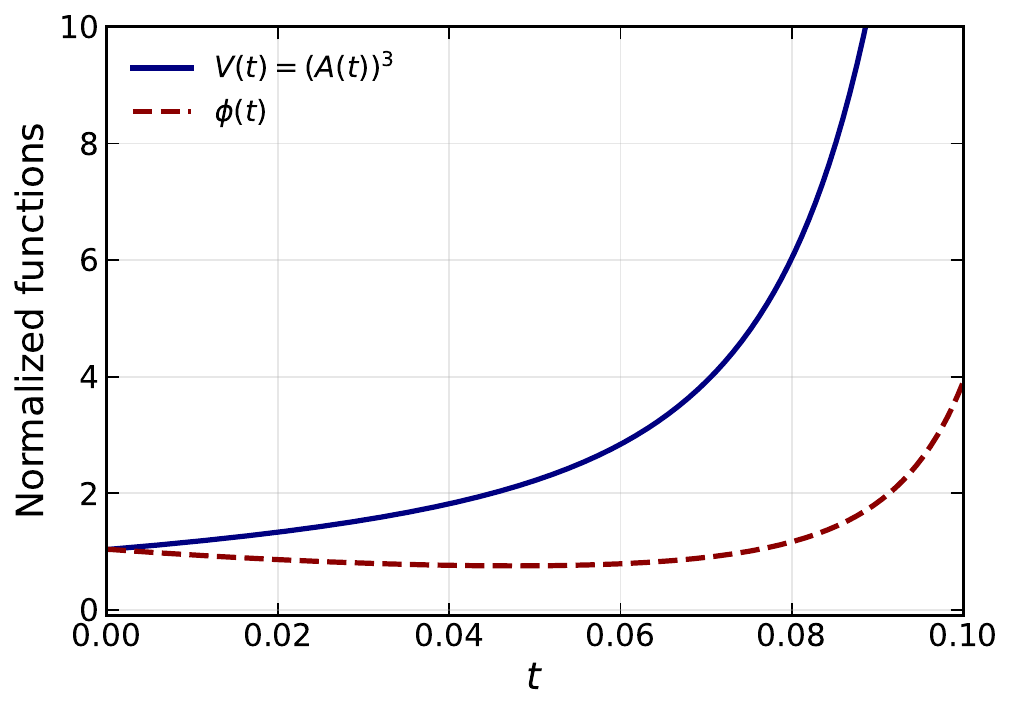}
 
\caption{Evolution of the normalized volume function,
$V(t)=\left[A(t)/A_0\right]^3$, and the normalized scalar field,
$\phi(t)/\phi_0$, corresponding to the exact solution given by
Eqs.~(\ref{scale-factor}) and (\ref{scalar-0}) for the quintessence branch with
the inverse power-law potential $V(\phi)\propto\phi^{-\sqrt{3}}$. The
plots were obtained for $\lambda=\sqrt{3}$, $p_0=1$, and $p_1=10$. The
solution exhibits a monotonically expanding universe, while the scalar
field undergoes an initial decrease followed by a continuous growth,
reflecting the interplay between the exponential and self-interaction
terms in the exact solution.}

\label{volumen-a0}
\end{center}
\end{figure}

Finally, the deceleration parameter is reconstructed from the exact scale
factor given in Eq.~(\ref{scale-factor}). The resulting evolution is shown in
Fig.~\ref{q-quinte-a0}. The solution begins in a strongly accelerated
regime, characterized by $q<-1$, reaches a minimum at early times, and
then evolves monotonically toward the asymptotic value $q=-1$. This
behavior indicates that the transient super-accelerated expansion is
gradually replaced by a de Sitter phase, which acts as the late-time
attractor of the cosmological solution.
\begin{figure}[h!]
\begin{center}
%\captionsetup{width=.8\textwidth}
\includegraphics[scale=0.5]{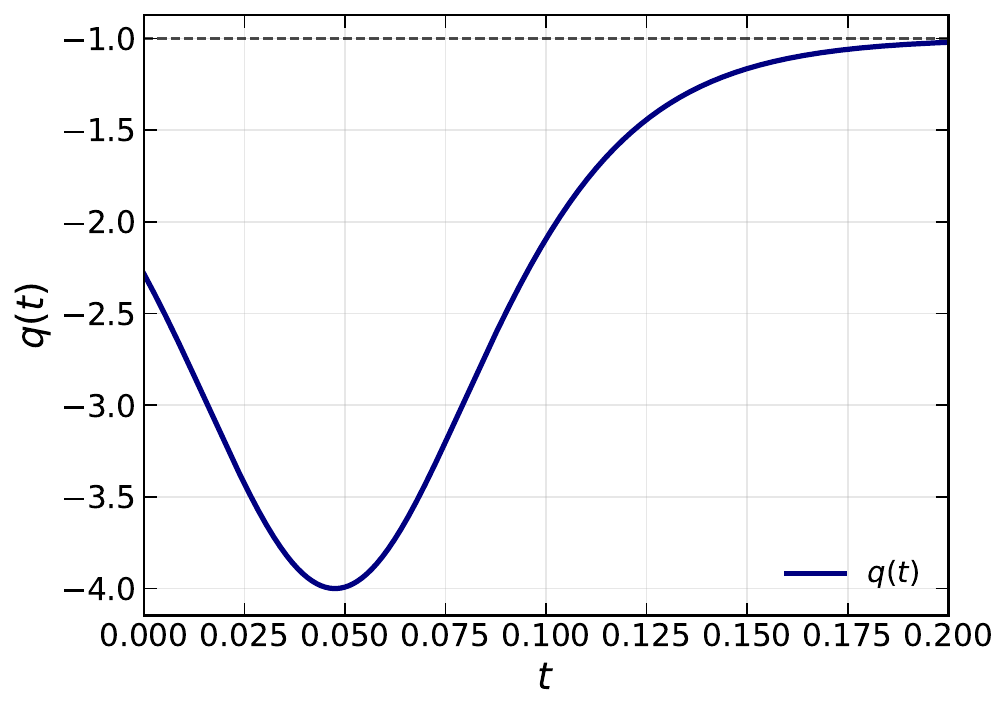}
\caption{Evolution of the deceleration parameter $q(t)$ reconstructed
from the exact scale factor given by Eq.~(\ref{scale-factor}). The same
parameter values used in Fig.~\ref{volumen-a0} are adopted, namely
$\lambda=\sqrt{3}$, $p_0=1$, and $p_1=10$. The deceleration parameter
remains below $-1$ throughout the evolution and asymptotically
approaches the de Sitter value $q=-1$, indicating a transition toward a
late-time accelerated expansion.} \label{q-quinte-a0}
\end{center}
\end{figure}

\newpage
\item{} $a>0$ imply that $\lambda < \sqrt{3}$, the master equation
(\ref{master}) can be written as
\begin{equation}
\int\frac{d\Pi_\Omega}{\Pi_\Omega^2+B\Pi_\Omega-C}= a\int dt
\nonumber
\end{equation}
where  $B=\frac{b}{ a}$ and $C=\frac{c}{ a}$, who solution become
\begin{equation}
\Pi_\Omega=-\frac{6\lambda p_0}{3-\lambda^2} - \frac{6\sqrt{3}
p_0}{3-\lambda^2} Coth\left(12\sqrt{3} p_0 \Delta t\right),
\end{equation}
thus the momenta to the scalar field (\ref{ecPiphi})
\begin{equation}
\Pi_\phi=\frac{3p_0}{3-\lambda^2}+ \frac{\sqrt{3} \lambda
p_0}{3-\lambda^2} Coth\left(12\sqrt{3} p_0 \Delta t\right),
\end{equation} and the solutions to $(\Omega,\varphi)$ become
\begin{eqnarray}
\Omega(t)=&=&\Omega_0 - \frac{12p_0 \lambda}{3-\lambda^2}\Delta
t-\frac{1}{3-\lambda^2}Ln\left[Sinh\left(12\sqrt{3} p_0 \Delta
t\right)  \right],
\\
\varphi(t)&=&\varphi_0 -\frac{72p_0}{3-\lambda^2} \Delta t
-\frac{2\lambda}{3-\lambda^2} Ln\left[Sinh\left((12\sqrt{3}p_0\,
\Delta t\right)\right].
\end{eqnarray}
so, the scale factor and the scalar field $\phi$ are
\begin{eqnarray}
A(t)&=&A_0 Csch^{\frac{1}{3-\lambda^2}} \left(12\sqrt{3} p_0 \Delta
t\right)\, Exp\left[- \frac{12p_0 \lambda}{3-\lambda^2}\Delta t
\right],\label{scale-a1}\\
\phi(t)&=&\phi_0\, Exp\left[-\frac{72p_0}{3-\lambda^2}\Delta t
\right]\, Csch^{\frac{2\lambda}{3-\lambda^2}}\left(12\sqrt{3}p_0
\Delta t\right), \label{phi-a1}
\end{eqnarray}
Figure~\ref{volumen-a1} shows the evolution of the normalized volume
function and the scalar field for the parameter choice
$\lambda=1/2$, corresponding to the inverse power-law potential
$V(\phi)=V_0\phi^{-1/2}$. In contrast to the previous solution, the
expansion develops more gradually, with both the volume function and
the scalar field increasing monotonically throughout the evolution.
This smoother behavior provides greater flexibility in adjusting the
model parameters to obtain a desired number of inflationary e-folds,
making the solution compatible with conventional inflationary
requirements \cite{rafael}.

\begin{figure}[h!]
\begin{center}
%\captionsetup{width=.8\textwidth}
\includegraphics[scale=0.5]{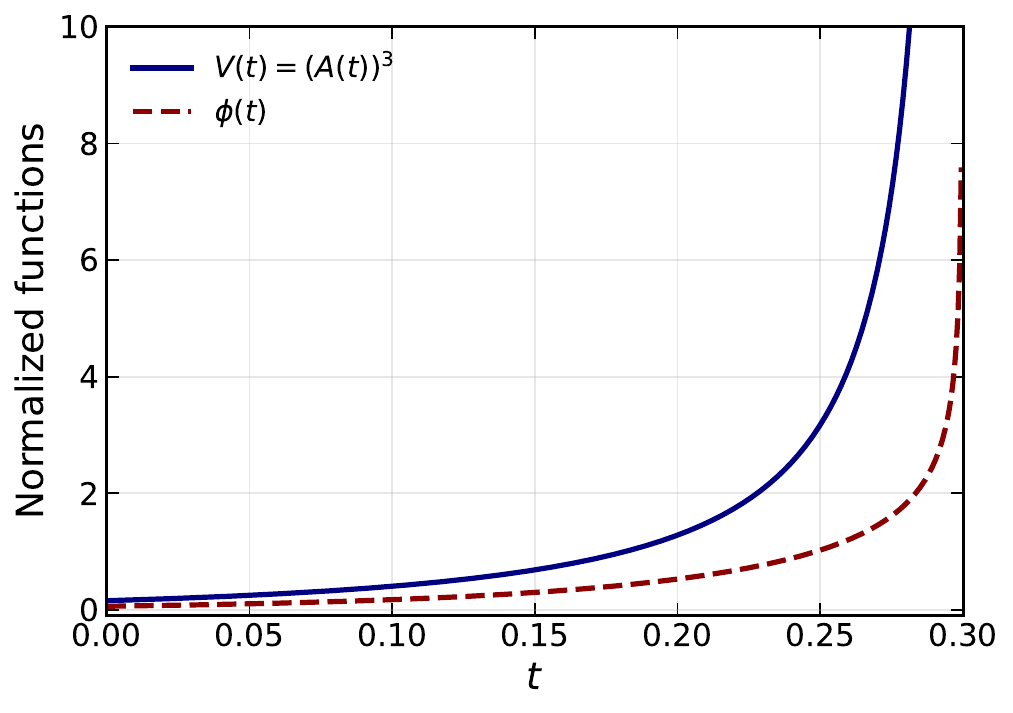}

\caption{Evolution of the normalized volume function,
$V(t)=\left[A(t)/A_0\right]^3$, and the normalized scalar field,
$\phi(t)/\phi_0$, corresponding to the exact solution given by
Eqs.~(\ref{scale-a1}) and (\ref{phi-a1}) for the quintessence branch
with the inverse power-law potential
$V(\phi)\propto\phi^{-1/2}$. The plots were obtained for
$\lambda=1/2$, $p_0=-0.3$, and $p_1=0.3$. Both the effective volume and
the scalar field increase monotonically, although with a significantly
smoother evolution than in the previous case, allowing for a more
gradual expansion of the universe.}
\label{volumen-a1}
\end{center}
\end{figure}
\newpage
\item{} $a<0$ imply  $\lambda > \sqrt{3}$ and we define $\tilde a=2(\lambda^2-3)$ the master equation become now
\begin{equation}
\int\frac{d\Pi_\Omega}{\Pi_\Omega^2+B\Pi_\Omega-C}=\tilde a\int dt
\nonumber
\end{equation} with $B=\frac{b}{\tilde a}$ and
$C=\frac{c}{\tilde a}$: who solution is
\begin{equation}
\Pi_\Omega=-\frac{6\lambda p_0}{\lambda^2-3} -
\frac{6\sqrt{2\lambda^2-3} p_0}{\lambda^2-3}
Coth\left(12\sqrt{2\lambda^2 -3} p_0 \Delta t\right)
\end{equation}

and the other momenta
\begin{equation}
\Pi_\varphi=\frac{p_0\lambda^2}{\lambda^2-3}+
\frac{\sqrt{2\lambda^2-3} \lambda p_0}{\lambda^2-3}
Coth\left(12\sqrt{2\lambda^2 3} p_0 \Delta t\right) + p_0
\end{equation}
and the solution to $(\Omega,\varphi)$ are
\begin{eqnarray}
\Omega(t)&=&\Omega_0 - \frac{12p_0 \lambda}{\lambda^2-3}\Delta
t-\frac{1}{\lambda^2-3}Ln\left[Sinh\left(12\sqrt{2\lambda^2-3} p_0
\Delta t\right)  \right], \\
\varphi(t)&=& \varphi_0
-\frac{24p_0(2\lambda^2-3)}{\lambda^2-3}t-\frac{2\lambda}{\lambda^2-3}Ln\left[Sinh\left(12\sqrt{2\lambda^2-3}p_0
\Delta t\right)\right].
\end{eqnarray}
and the scale factor and the function $\phi$ become
\begin{eqnarray}
A(t)&=&A_0 Csch^{\frac{1}{\lambda^2-3}} \left(12\sqrt{2\lambda^2-3}
p_0 \Delta t\right)\, Exp\left[- \frac{12p_0
\lambda}{\lambda^2-3}\Delta t \right]. \label{scale-mayor} \\
\phi(t)&=& \phi_0 Exp\left[
-\frac{24p_0(2\lambda^2-3)}{\lambda^2-3}t\right]\,Csch^{\frac{2\lambda}{\lambda^2-3}}\left(12\sqrt{2\lambda^2-3}p_0
\Delta t\right). \label{phi-mayor}
\end{eqnarray}

Figure~\ref{volumen-mayor} displays the evolution of the normalized
volume function and the scalar field for the representative choice
$\lambda=4$, corresponding to the potential
$V(\phi)=V_0\phi^{-4}$. In contrast to the previous cases, the volume
function grows only slowly over most of the evolution, while the scalar
field remains nearly constant before increasing at late times. For the
parameter values considered, the resulting expansion is not sufficiently
rapid to sustain an inflationary phase with an adequate number of
e-folds. Consequently, this representative solution does not provide a
viable inflationary scenario, illustrating that the regime
$\lambda>\sqrt{3}$ leads to substantially weaker inflationary dynamics
than the cases discussed previously.
\begin{figure}[h]
\begin{center}
%\captionsetup{width=.8\textwidth}
\includegraphics[scale=0.5]{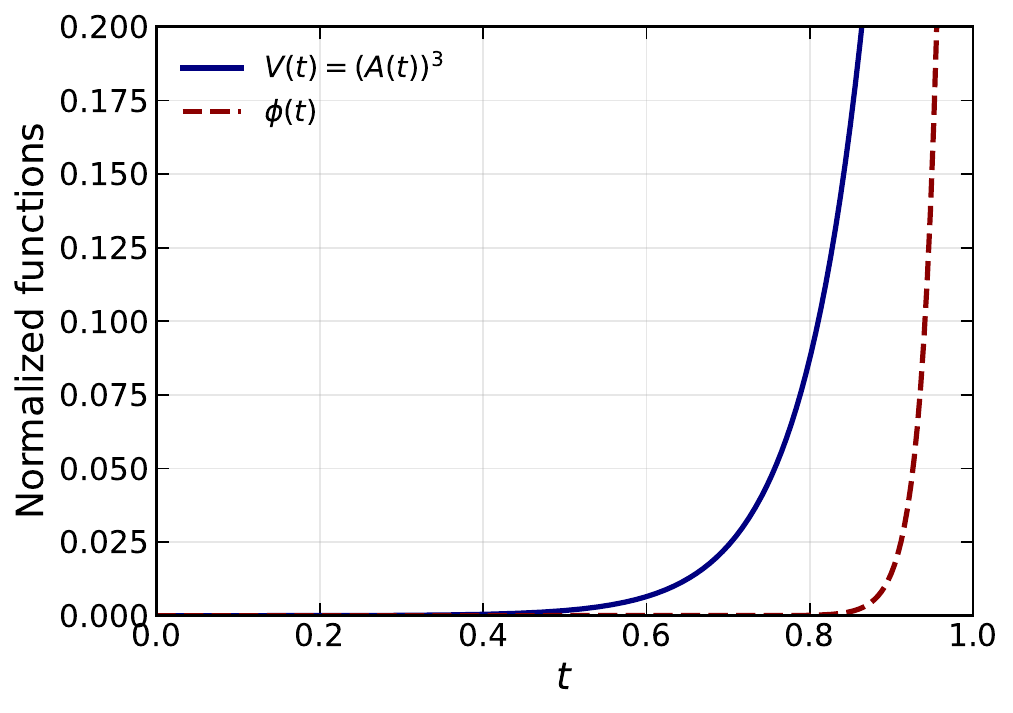}

\caption{Evolution of the normalized volume function,
$V(t)=\left[A(t)/A_0\right]^3$, and the normalized scalar field,
$\phi(t)/\phi_0$, corresponding to the exact solution given by
Eqs.~(\ref{scale-mayor}) and (\ref{phi-mayor}) for the quintessence
branch with the inverse power-law potential
$V(\phi)\propto\phi^{-4}$. The curves were obtained for
$\lambda=4$, $p_0=-1/2$, and $p_1=1$. In comparison with the previous
cases, both the effective volume and the scalar field evolve much more
slowly, leading to a significantly weaker expansion during the
interval considered.} \label{volumen-mayor}
\end{center}
\end{figure}

Interestingly, this behavior is qualitatively consistent with the
quantum analysis presented in Sec \ref{qap}, where the corresponding branch
is associated with a comparatively lower probability density. Although
this observation should not be interpreted as a direct physical
connection, it suggests an intriguing correspondence between the
classical dynamics and the quantum solutions.

\end{enumerate}

\subsection{Phantom field, $\epsilon=-1$}
The corresponding Hamiltonian density
\begin{equation}
\rm {\cal H}= \frac{e^{-3\Omega}}{24} \left[ \Pi_\Omega^2+12
\Pi_\varphi^2-24  V_0 e^{6\Omega-\lambda \varphi}\right].
\label{hamifrw-phantom}
\end{equation}
In the gauge $\rm N=24 e^{3\Omega}$ and using the Hamilton approach,
we have the following set of equation
\begin{eqnarray}
\rm \dot \Omega&=&\rm 2 \Pi_\Omega, \qquad \dot \varphi= 24 \Pi_\varphi, \nonumber\\
\rm \dot \Pi_\Omega&=& \rm 144 V_0 e^{6\Omega- \lambda \varphi},
\qquad \dot \Pi_\phi= -24\lambda V_0 e^{6\Omega- \lambda
\varphi},\label{new-var}
\end{eqnarray}
from (\ref{new-var}) we find the relation between the momenta
\begin{equation}
\Pi_\varphi=-\frac{\lambda\Pi_\Omega}{6}+p_0
\end{equation}
that inserting in the equation for the scalar field
\begin{equation}
\varphi=-2\lambda\Omega+24p_0t+\alpha_0. \label{phi-phantom}
\end{equation}

\begin{figure}[ht!]
\begin{center}
%\captionsetup{width=.8\textwidth}
%\includegraphics[scale=0.6]{scale-pha.pdf}
%\includegraphics[scale=0.6]{phi-pha.pdf}
\includegraphics[scale=0.9]{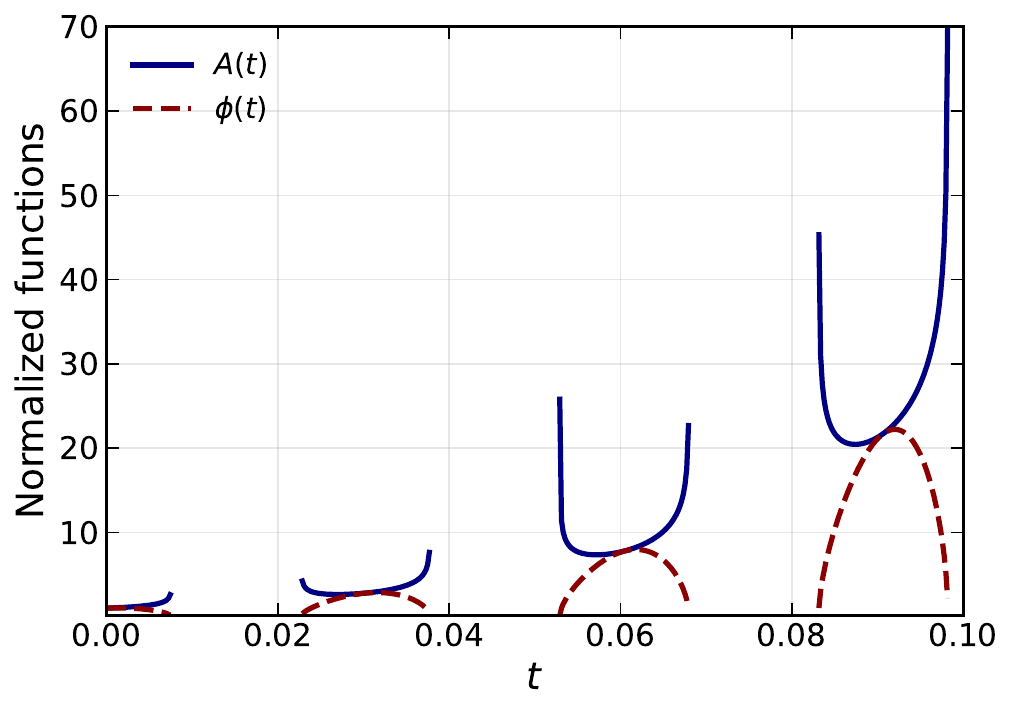}

\caption{Evolution of the normalized scale factor,
$A(t)/A_0$, and the normalized scalar field,
$\phi(t)/\phi_0$, corresponding to the exact phantom solution given by
Eqs.~(\ref{scale-phantom}) and (\ref{phi-phantom}) with the inverse
power-law potential $V(\phi)\propto\phi^{-\lambda}$. The curves are
shown for $p_0=10$, $p_1=1$, while the solution is valid for arbitrary
values of $\lambda$. The evolution consists of successive regular
branches separated by singular points, with the scalar field and the
scale factor exhibiting complementary behavior within each branch.
}\label{volumen-pha}
\end{center}
\end{figure}

Employing the hamiltonian constraint, we find the following master
equation
\begin{equation}
\frac{d\Pi_\Omega}{\left(\Pi_\Omega- \frac{6\lambda
p_0}{3+\lambda^2}\right)^2+ \left(\frac{6\sqrt{3}p_0}{3+\lambda^2}
\right)^2} =2(3+\lambda^2)dt, \label{master-phantom}
\end{equation}
who solution for the momenta $\Pi_\Omega$ is
\begin{equation}
\Pi_\Omega(t)=\frac{6\lambda p_0}{3+\lambda^2} + \frac{6\sqrt{3}
p_0}{3+\lambda^2}\, Tan \left(12\sqrt{3} p_0 \Delta t \right),
\label{mom}
\end{equation}
so, the other temporal variables become
\begin{eqnarray}
\Omega(t)&=&\Omega_0 +\frac{12\lambda p_0}{3+\lambda^2}\Delta t -
\frac{1}{3+\lambda^2}\, Ln\left[Cos\left(12\sqrt{3}p_0 \Delta t
\right) \right], \label{ome-pha}\\
\varphi(t)&=&\varphi_0+\frac{72p_0}{3+\lambda^2}\Delta
t+\frac{2\lambda}{3+\lambda^2}\, Ln\left[Cos\left(12\sqrt{3}p_0
\Delta t \right) \right], \label{phi-pha}
\end{eqnarray}
thus, the scale factor and the scalar field $\phi$ are
\begin{eqnarray}
A(t)&=&A_0 \, e^{\frac{12\lambda p_0}{3+\lambda^2}(t-p1)}\,
Sec^{\frac{1}{3+\lambda^2}} \left(12\sqrt{3}p_0 \Delta t \right)
,\label{scale-phantom}\\
\phi(t)&=& \phi_0 e^{\frac{12\lambda p_0}{3+\lambda^2}(t-p1)}\,
Cos^{\frac{2\lambda}{3+\lambda^2}} \left(12\sqrt{3}p_0 \Delta t
\right). \label{phi-phantom}
\end{eqnarray}
Figure~\ref{volumen-pha} illustrates the evolution of the normalized
scale factor and the scalar field in the phantom sector. The exact
solution is valid for arbitrary values of the parameter $\lambda$ and
exhibits a sequence of disconnected branches separated by singular
points. Within each branch, the scale factor grows monotonically,
whereas the scalar field displays an opposite trend, reaching its
largest values when the expansion rate is comparatively weaker and
decreasing as the scale factor grows. This complementary behavior
suggests that the scalar field continuously drives the cosmological
evolution while the dynamics repeat from one branch to the next.

\begin{figure}[ht!]
\begin{center}
%\captionsetup{width=.8\textwidth}
\includegraphics[scale=0.8]{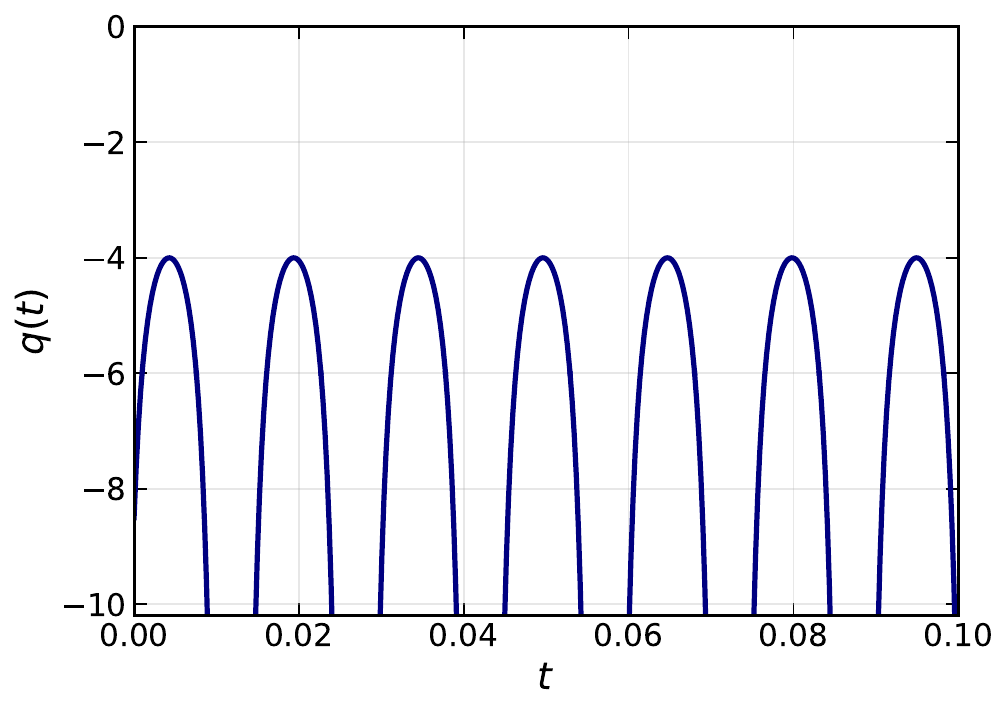}
\caption{Evolution of the deceleration parameter reconstructed from the
exact phantom scale factor given by Eq.~(\ref{scale-phantom}). The same
parameter values adopted in Fig.~\ref{volumen-pha} are used. The
solution consists of a sequence of disconnected branches separated by
discontinuities, while within each branch the deceleration parameter
remains below $-1$, indicating a recurrent super-accelerated
cosmological evolution.} \label{q-phantom}
\end{center}
\end{figure}

Finally, the deceleration parameter, defined by
Eq.~(\ref{q-deceleration}), is reconstructed from the exact phantom
scale factor given by Eq.~(\ref{scale-phantom}). The corresponding
evolution is shown in Fig.~\ref{q-phantom}. In agreement with the
behavior of the scale factor, the deceleration parameter is composed of
a sequence of disconnected branches separated by discontinuities.
Within each branch, the universe remains in a super-accelerated regime
($q<-1$), while the periodic repetition of the branches reflects the
cyclic character of the exact phantom solution. Interestingly, this
behavior is qualitatively consistent with the quantum analysis
presented in Sec. \ref{qap}, where the corresponding branch is associated
with a comparatively lower probability density. Although this
observation does not establish a direct connection between the
classical and quantum descriptions, it suggests an intriguing
correspondence between the two pictures.

This scheme can be generalized to multi-fields, but as shown in reference \cite{CQG-2021}, this is not possible in a standard scheme, 
rather it is necessary to introduce a free parameter in a mixed term of the kinetic energy of scalar fields, see reference \cite{bianchi-i}
 where the formalism was applied to the anisotropic cosmological Bianchi type I.

%%%%%%%
%%%%%%%

\section{quantum approach\label{qap}}
On the Wheeler-DeWitt (WDW) equation there are a lot of papers
dealing with different problems, for example in \cite{Gibbons}, they
asked the question of what a typical wave function for the universe
is. On the other hand, the best candidates for quantum solutions
become those that have a damping behavior with respect to the scale
factor, since these allow to obtain good classical solutions when
using the WKB approximation for any scenario in the evolution of our
universe \cite{HH,H}.

 The Wheeler-DeWitt equation   for
this model is acquired by replacing
 $\rm \Pi_{q^\mu}=-i\hbar \partial_{q^\mu}$ in (\ref {hamifrw}).
 The factor $\rm e^{-3\Omega}$ may be factor ordered with $\rm \hat \Pi_\Omega$ in many ways. Hartle and
Hawking \citep{HH} have suggested what might be called a
semi-general factor ordering, which in this case would order $\rm
e^{-3\Omega} \hat \Pi^2_\Omega$ as
\begin{eqnarray}
\rm - e^{-(3- Q)\Omega}\, \partial_\Omega e^{-Q\Omega}
\partial_\Omega&=&\rm - e^{-3\Omega}\, \partial^2_\Omega + Q\,
e^{-3\Omega} \partial_\Omega, \label {hh}
\end{eqnarray}
where  Q is any real constant that measure the ambiguity in the
factor ordering for the variable $\Omega$. In the following we will
assume such factor ordering for the Wheeler-DeWitt equation, which
becomes
\subsection{Quintessence case, $\epsilon=1$}
The Wheeler-DeWitt equation is written as
\begin{equation}
\rm \hbar^2 \Box \Psi+ \hbar^2 Q\frac{\partial \Psi}{\partial
\Omega}- e^{6\Omega}U(\varphi)\Psi=0, \label{wdwmod}
\end{equation}
where  $\rm \Box=-\frac{\partial^2}{\partial
\Omega^2}+12\frac{\partial^2}{\partial \varphi^2}$ is the
d'Alambertian in the coordinates $q^\mu=(\Omega,\varphi)$ and the
potential is $\rm U=  +24V_0 e^{-\lambda \varphi} $.

Making the canonical transformation
\begin{equation}
\rm x=6\Omega-\lambda \varphi, \qquad y=\alpha_1 \Omega + \alpha_2
\varphi, \label{can-trans}
\end{equation}
Eq. (\ref{wdwmod}) is rewritten as
\begin{eqnarray}
&&12\left(\lambda^2-3\right)\frac{\partial^2 \Psi(x,y)}{\partial
x^2}+ \left(12\alpha_2^2-\alpha_1^2 \right)\frac{\partial^2
\Psi(x,y)}{\partial y^2}-  12\left(\alpha_1+2\lambda \alpha_2
\right) \frac{\partial^2}{\partial x
\partial y}\Psi(x,y) \nonumber\\
&&+  6Q\left(\frac{\partial}{\partial x} +
\alpha_1\frac{\partial}{\partial y} \right)\Psi(x,y)-\frac{24
V_0}{\hbar^2} e^{x}\Psi=0
\end{eqnarray}

 This equation have the following solution $\rm
\Psi(x,y)=e^y \, e^{u(x)}$, where the function $\rm u(x)$ satisfy
the ordinary differential equation
\begin{equation}
12\left(\lambda^2-3\right)\frac{d^2 u(x)}{d
x^2}-\beta_3\frac{du(x)}{dx}+ \left(\beta_1-\beta_2 e^x\right)u(x)=0
\end{equation}
where $\beta_1=12\alpha_2^2-\alpha_1^2+6Q\alpha_1$,
$\beta_2=24\frac{V_0}{\hbar^2}$ and $\beta_3=12\alpha_1+24\lambda
\alpha_2-6Q$. In the following, this equation we solve for subcases.

\begin{enumerate}
\item{} When $\lambda=\sqrt{3}$, the wave function is
$$\rm \Psi(x,y)=\Psi_0 e^y Exp\left[\frac{\beta_1}{\beta_3}x - \frac{\beta_2}{\beta_3}e^x \right],$$
that written in the original variables $(A,\phi)$ is
\begin{equation}
 \Psi(A,\phi)=\Psi_0 A^{\nu_1} \phi^{\nu_2}\, Exp\left(-
\frac{\beta_2}{\beta_3} A^6 \phi^{-\sqrt{3}} \right], \label{raiz3}
\end{equation}
with the constants $\beta_1$ and $\beta_2$ are the same, but
$\beta_3=6(2\alpha_1+4\sqrt{3}\alpha_2-Q)$ and the constants $\rm
\nu_1=\alpha_1+6\frac{\beta_1}{\beta_3}$ and $\nu_2 =\alpha_2-
\sqrt{3}\frac{\beta_1}{\beta_3}$,
 and chosing the values to constants
$\alpha_1=-2\sqrt{3}\alpha_2$, $\alpha_2=2$ and $Q=-2$, the behavior
of the probability density of the wave function appear in the figure
\ref{lambda3}

\begin{figure}[ht!]
\begin{center}
\includegraphics[scale=0.8]{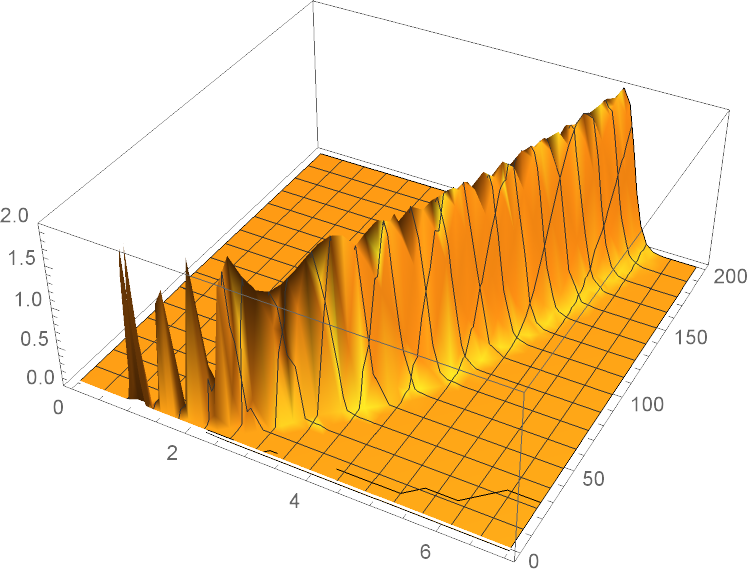}
\caption{the probability density of the equation (\ref{raiz3}),
where we choose the values $\Psi_0=1$, $Q=-2$, $\alpha_2=2$ and
$\alpha_1=-2\sqrt{3}\alpha_2$. } \label{lambda3}
\end{center}
\end{figure}
This behavior holds true for other values of $Q<0$, but $\Psi_0$
must be manually adjusted to achieve similar behavior. At Q=0 it
becomes indeterminate, but for $Q>0$ it reverts to this behavior,
although the wave function grows rapidly, requiring $\Psi_0$ to be
made increasingly smaller.

For other values of $\alpha_1$ and $\alpha_2$, a similar behavior is
maintained, but $\Psi_0$ must be small, since the wave function is
non-normalized, as shown in figure \ref{lambda4}.
\begin{figure}[ht!]
\begin{center}
\includegraphics[scale=0.8]{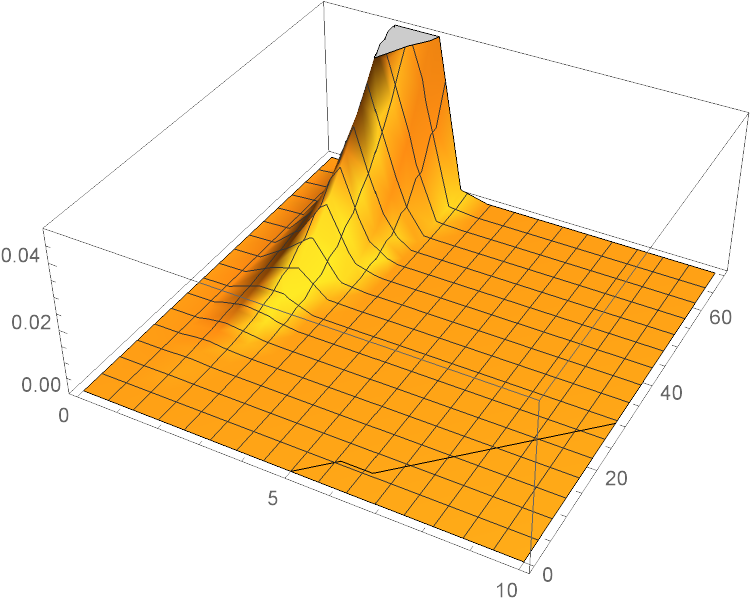}
\caption{the probability density of the equation (\ref{raiz3}),
where we choose the values $\Psi_0=10^{-6}$, $Q=2$, $\alpha_1=1$
$\alpha_2=2$. } \label{lambda4}
\end{center}
\end{figure}
In both figures it can be observed that the evolution of the wave
function increases in the direction of the scalar field $\phi$, and
the wave function present a spontaneous collapse direction of the
scale factor A, essential for the classical universe.  So the scalar
field will have a very noticeable presence in the classical behavior
in its temporal evolution.

\item{} For $\lambda > \sqrt{3}$, the differential equation for the
function $u(x)$ become
\begin{equation}
\frac{d^2 u(x)}{d x^2}-\eta_1\frac{du(x)}{dx}+ \left(\eta_2-\eta_3
e^x\right)u(x)=0
\end{equation}
where the constants $\eta_1=\frac{2\alpha_1+4\lambda
\alpha_2-Q}{2\left(\lambda^2-3\right)}$,
$\eta_2=\frac{12\alpha_2^2-\alpha_1^2+6Q\alpha_1}{12\left(\lambda^2-3\right)}$
and $\eta_3=\frac{2V_0}{\hbar^2\left(\lambda^2-3\right)}$, following
at Polyanin \cite{polyanin}, the solution become
\begin{equation}
u(x)=u_0\, e^{\frac{\eta_1}{2}x}
K_\nu\left(2\sqrt{\eta_3}e^{\frac{x}{2}}\right), \label{sol-mayor}
\end{equation}
where the order  $\nu=\sqrt{\eta_1^2-4\eta_2}$ and $K_\nu$ is the
modified Bessel function. Then the complete solution for this case
become
\begin{eqnarray} \Psi(x,y)&=&u_0 e^{\frac{\eta_1}{2}x+y}\,
K_\nu\left(2\sqrt{\eta_3}e^{\frac{x}{2}}\right)\nonumber\\
&=&\Psi_0 A^{\alpha_1+3\eta_1}\,\phi^{\alpha_2-\frac{\lambda
\eta_1}{2}}\, K_\nu\left(2\sqrt{\eta_3}A^3
\,\phi^{-\frac{\lambda}{2}} \right).\label{psi-mayor}
\end{eqnarray}

In the figure \ref{mayor3} it can be observed that the parameter Q
serves to shift the probability density to the left $(Q>0)$ or to
the right $(Q<0)$ in the direction of the scale factor evolution,
but also so that the probability density has larger values $(Q>>1)$
and
 viceversa, decreases in Q close to $-1$, see figure
\ref{mayor33}. However, the probability density is very small, which
may make a classic scenario with $\lambda^2 > 3$ unlikely.
\begin{figure}[ht!]
\begin{center}
\includegraphics[scale=0.6]{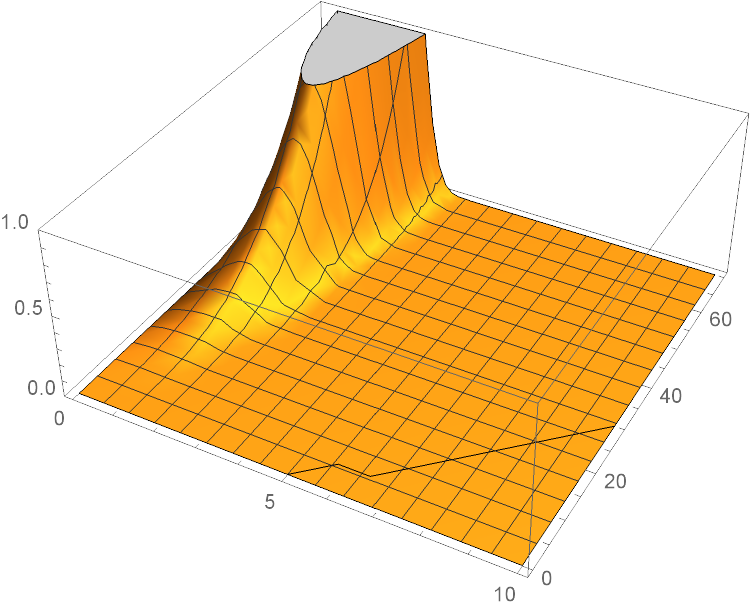}
\includegraphics[scale=0.6]{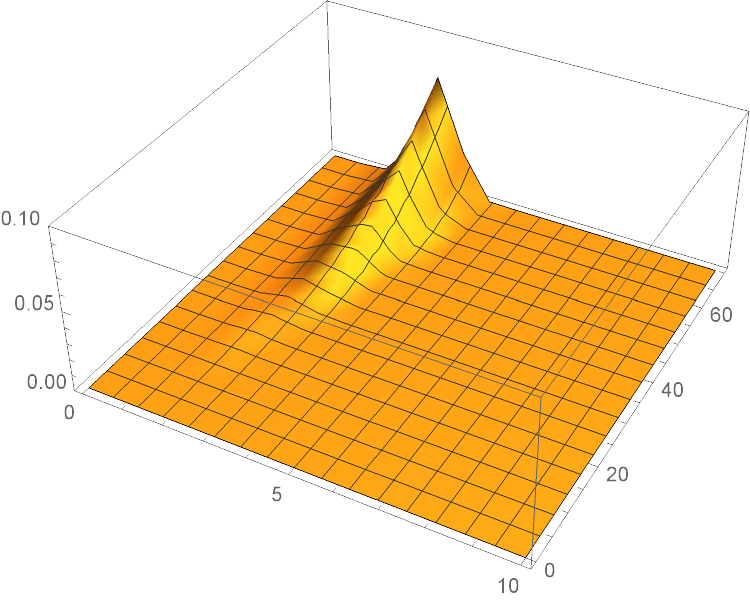}
\caption{the probability density of the equation (\ref{psi-mayor}),
where we choose the values $\lambda=2$,$\Psi_0=1$, $\alpha_2=2$,
$V_0=10$ for both figures, but $Q=6$ for the left and $Q=-8$ for the
right side . } \label{mayor3}
\end{center}
\end{figure}

\begin{figure}[ht!]
\begin{center}

\includegraphics[scale=1]{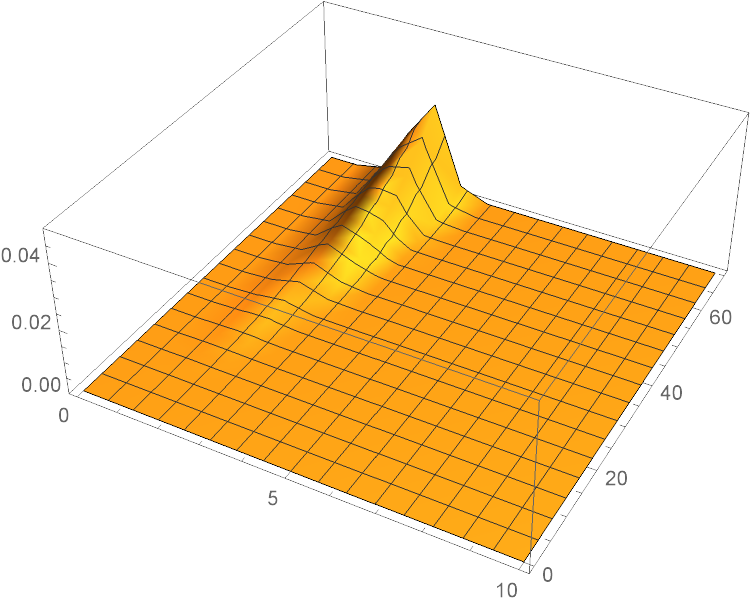}
\caption{the probability density of the equation (\ref{psi-mayor}),
where we choose the values $\lambda=2$,$\Psi_0=1$, $\alpha_2=2$,
$V_0=10$ and $Q=-2$. } \label{mayor33}
\end{center}
\end{figure}

\item{}  For $\lambda < \sqrt{3}$, the differential equation for the
function $u(x)$ become
\begin{equation}
\frac{d^2 u(x)}{d x^2}+\eta_1\frac{du(x)}{dx}+ \left(-\eta_2+\eta_3
e^x\right)u(x)=0
\end{equation}
where the constants $\eta_1=\frac{2\alpha_1+4\lambda
\alpha_2-Q}{2\left(3-\lambda^2\right)}$,
$\eta_2=\frac{12\alpha_2^2-\alpha_1^2+6Q\alpha_1}{12\left(3-\lambda^2\right)}$
and $\eta_3=\frac{2V_0}{\hbar^2\left(3-\lambda^2\right)}$, following
at Polyanin \cite{polyanin}, the solution become
\begin{equation}
u(x)=u_0\, e^{-\frac{\eta_1}{2}x}
K_\nu\left(2\sqrt{\eta_3}e^{\frac{x}{2}}\right), \label{sol-mayor}
\end{equation}
where the order  $\nu=\sqrt{\eta_1^2+4\eta_2}$ and $J_\nu$ is the
ordinary Bessel function. Then the complete solution for this case
become
\begin{eqnarray} \Psi(x,y)&=&u_0 e^{-\frac{\eta_1}{2}x+y}\,
J_\nu\left(2\sqrt{\eta_3}e^{\frac{x}{2}}\right)\nonumber\\
&=&\Psi_0 A^{\alpha_1-3\eta_1}\,\phi^{\alpha_2+\frac{\lambda
\eta_1}{2}}\, J_\nu\left(2\sqrt{\eta_3}A^3
\,\phi^{-\frac{\lambda}{2}} \right).\label{psi-menor}
\end{eqnarray}

In the figure \ref{menor3}, the graphical analysis shows that when
the parameter $Q \in [1.36, 1.37)$, the order of the ordinary Bessel
function becomes imaginary, requiring a different approach to the
solution, constructing a wave packet in the order of the function.
Furthermore, this quantum universe is more likely to produce a
classical inflationary universe, given that the wave function is
increasing for other values of this parameter, and  the $\lambda$
parameter should possibly be smaller than one, following the ideas
of the reference \cite{rafael}.

\begin{figure}[ht!]
\begin{center}
\includegraphics[scale=0.6]{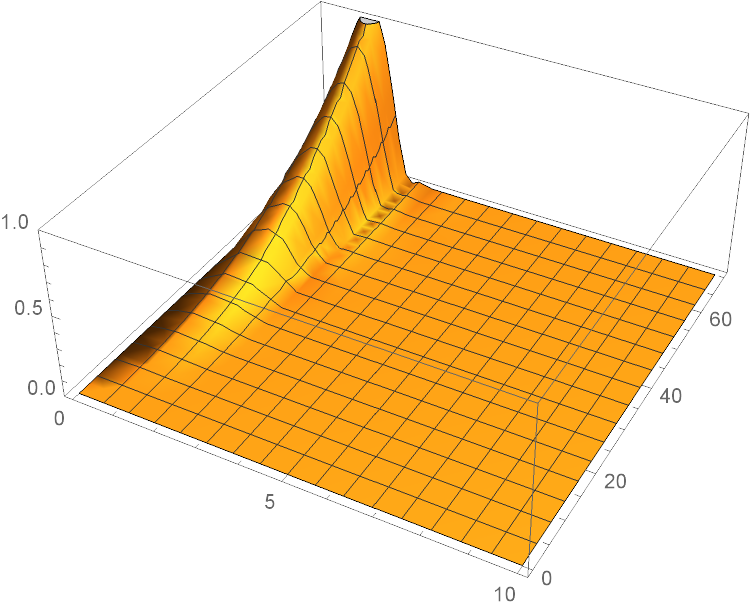}
\includegraphics[scale=0.6]{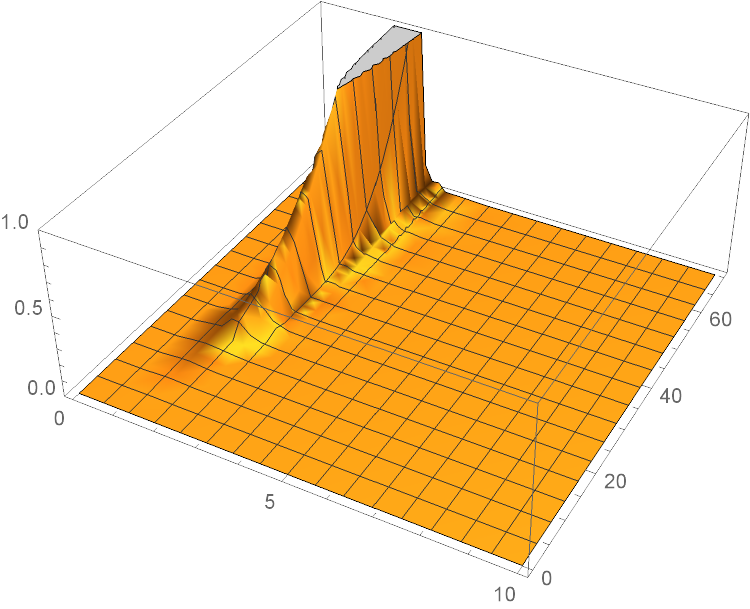}
\caption{the probability density of the equation (\ref{psi-menor}),
where we choose the values $\lambda=1$, $\alpha_2=2$,
$\alpha_1=-2\lambda \alpha_2$ and $V_0=1$ for both figures, but
$Q=2$  and $\Psi_0=10^{-\frac{5}{2}}$ for the left and $Q=-2$ and
$\Psi_0=10^{-\frac{3}{2}}$for the right side . } \label{menor3}
\end{center}
\end{figure}

\end{enumerate}
%%%%%%%
%%%%%%%
\subsection{Phantom case, $\epsilon=-1$}
The Wheeler-DeWitt equation is written as
\begin{equation}
\rm \hbar^2 \Box \Psi+ \hbar^2 Q\frac{\partial \Psi}{\partial
\Omega}- e^{6\Omega}U(\varphi)\Psi=0, \label{wdwmod}
\end{equation}
where  $\rm \Box=-\frac{\partial^2}{\partial
\Omega^2}-12\frac{\partial^2}{\partial \varphi^2}$ is the
d'Alambertian in the coordinates $q^\mu=(\Omega,\varphi)$ and the
potential is $\rm U=  +24V_0 e^{-\lambda \varphi} $.

Making the canonical transformation
\begin{equation}
\rm x=6\Omega-\lambda \varphi, \qquad y=\alpha_1 \Omega + \alpha_2
\varphi, \label{can-trans}
\end{equation}
Eq. (\ref{wdwmod}) is rewritten as
\begin{eqnarray}
&&-12\left(\lambda^2+3\right)\frac{\partial^2 \Psi(x,y)}{\partial
x^2}- \left(12\alpha_2^2+\alpha_1^2 \right)\frac{\partial^2
\Psi(x,y)}{\partial y^2}-  12\left(\alpha_1-2\lambda \alpha_2
\right) \frac{\partial^2}{\partial x
\partial y}\Psi(x,y) \nonumber\\
&&+  6Q\left(\frac{\partial}{\partial x} +
\alpha_1\frac{\partial}{\partial y} \right)\Psi(x,y)-\frac{24
V_0}{\hbar^2} e^{x}\Psi=0
\end{eqnarray}

 This equation have the following solution $\rm
\Psi(x,y)=e^y \, e^{u(x)}$, where the function $\rm u(x)$ satisfy
the ordinary differential equation
\begin{equation}
\frac{d^2 u(x)}{d x^2}-\beta_3\frac{du(x)}{dx}+
\left(\beta_1+\beta_2 e^x\right)u(x)=0
\end{equation}
where
$\beta_1=\frac{12\alpha_2^2+\alpha_1^2-6Q\alpha_1}{12(\lambda^2+3)}$,
$\beta_2=\frac{2V_0}{\hbar^2(\lambda^2+3)}$ and
$\beta_3=\frac{-2\alpha_1+4\lambda \alpha_2+Q}{2(\lambda^2+3)}$. The
general solution become
\begin{equation}
u(x)=u_0 e^{\frac{\beta_3}{2}x} \, J_\nu\left( 2\sqrt{\beta_2}
e^\frac{x}{2}\right) \label{phantom-solution}
\end{equation}
with the order $\nu=\sqrt{\beta^2-4\beta_1}$. Thus the wave function
in the original variable become
\begin{equation}
\Psi(A,\phi)=\Psi_0 A^{\theta_1} \phi^{\theta_2}\, J_\nu \left(A^3
\, \phi^{-\frac{\lambda }{2}} \right), \label{psi-phantom}
\end{equation}
where $\theta_1=\alpha_1+3\beta_3$ and
$\theta_2=\alpha_2-\frac{1}{2}\lambda \beta_3$.

\begin{figure}[ht!]
\begin{center}
\includegraphics[scale=0.6]{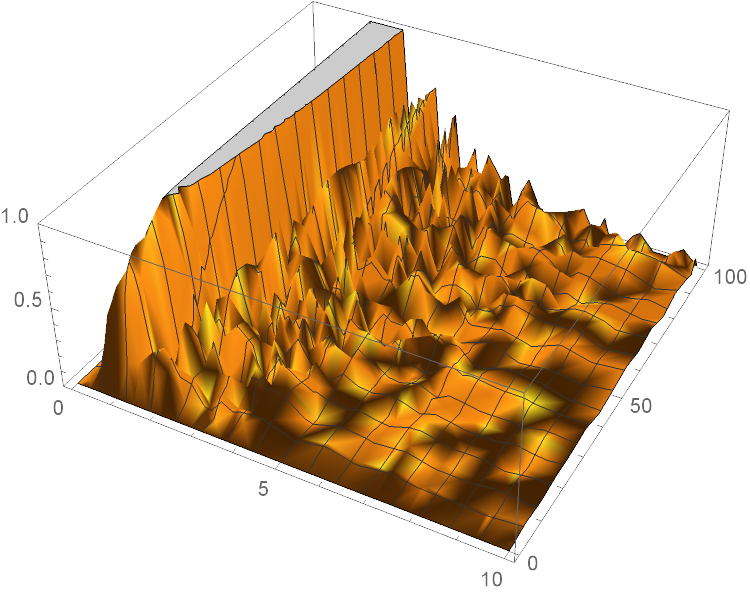}

\caption{the probability density of the equation
(\ref{psi-phantom}), where we choose the values $\lambda=1$,
$\alpha_2=2$, $\alpha_1=-2\lambda \alpha_2$ and $V_0=1$ for both
figures, but $Q=2$  and $\Psi_0=1$. } \label{phan-1}
\end{center}
\end{figure}

In the phantom scenario in this propose, the probability density
function is highly sensitive to the values of the constants,
exhibiting an essential universe and fuzzy copies in the evolution
of the scale factor. Figure \ref{phan-1} presents one such case
where for $\lambda= 1$, the constants $a_1=2 \lambda \alpha_2$,
$\alpha_2=0.1$, $V_0=1$, and Q=1. For other values, the order of the
ordinary Bessel function may be imaginary.

\section{Final remarks}

In this work we introduce a proposal into a combination of
Saez-Ballester-K-Essence like theory, with standard kinetic energy
term, to analyze the evolution of the universe in the two scenarios
introduced in the literature: quintessence and  phantom scalar
field, employing a power law scalar potential fields in cosmology $
\rm V(\phi)=V_0 \phi^{- \lambda}$.

 What we
can infer from these results is that the evolution of the universe
is controlled by this field when the scalar field is more
progressive than the evolution of the universe; subsequently, this
same fields should attenuate this evolution of the volume of the
universe, being present at all times as a cosmic background. This
behavior was observed in another alternative theory to general
relativity for  anisotropic cosmological Bianchi type I, with a
"geometric scalar field", where the evolution of the universe in its
different stages is modulated via this geometric scalar field
\cite{fr}. When $\lambda$ is very small, ($\lambda=\frac{1}{2}$),
the evolution is not so explosive and the free parameters can be
adjusted in such a way that we obtain the e-folds similar to the
reference \cite{rafael}, but not when $\lambda=\sqrt{3}$, the
evolution is very expansive giving an eternal inflation.

However, in the phantom scenario, the classic solution of
the scale factor shows a periodic scenario with a tendency that each
time it arises, its initial evolution becomes increasingly larger,
which would make the model not viable in this formalism, his trend
is shown in quantum space, where various diffuse universes appear
with lower probability.

From a quantum perspective, the probability density behaves as
expected in the evolution of the scale factor, it has a decaying
behavior, and the same pattern is observed in the evolution of the
scalar field, which we attribute to it remaining at the classical
level as a cosmic background in the evolution of the universe, as
presented in the cases studied.

Several subcases are possibly linked to the fact that in the space
of quantum solutions, the probability density shows a universe of
lower or higher probability, as can be seen in the section on
quantum solutions, which would dictate that subcase is more likely
to produce inflation in this formalism.

\acknowledgments{ \noindent This work was partially supported by
PROMEP grants UGTO-CA-3.  J.S. was partially supported SNI-CONACYT.
A.G. acknowledge support from SECIHTI post-doctoral fellowships.
Many calculations where done by Symbolic Program REDUCE 3.8.}

\end{document}